\documentclass{JAIS}

\journal{JAIS-ID}
\vol{20xx}

\received{xx January 2018}
\published{xx March 2018}

\def\be{\begin{equation}}
\def\ee{\end{equation}}
\def\bea{\begin{eqnarray}}
\def\eea{\end{eqnarray}}

\usepackage{caption}
\usepackage{lineno}
\usepackage{booktabs}
\usepackage{upgreek}
\usepackage{orcidlink} 
\usepackage{subcaption}
\usepackage{xcolor}
\usepackage{amsmath}
\usepackage{soul}

\articletype{}

\begin{document}

\title{Active interrogation of underground piezoelectric fabrics using high energy muon beams propagating across seismogenic faults}

\author{
L.~Serafini\,\orcidlink{0000-0002-4367-1555}\auno{1},
A.~Bacci\,\orcidlink{0000-0001-6010-9225}\auno{1},
L.~Bandiera\,\orcidlink{0000-0002-5537-9674}\auno{2},
F.~Broggi\,\orcidlink{0000-0003-1628-1604}\auno{1},
I.~Drebot\,\orcidlink{0000-0002-9152-0102}\auno{1},
A.~Frazzitta\,\orcidlink{0009-0009-3678-8497}\auno{1,3},
A.~M.~Marotta\,\orcidlink{0000-0002-9580-5981}\auno{4},
G.~Muttoni\,\orcidlink{0000-0001-7908-1664}\auno{4},
G.~Paternò\,\orcidlink{0000-0002-3022-3286}\auno{2},
V.~Petrillo\,\orcidlink{0000-0002-8556-3384}\auno{1,5},
M.~Rossetti~Conti\,\orcidlink{0000-0002-5767-3850}\auno{1},
A.~R.~Rossi\,\orcidlink{0000-0002-6216-8664}\auno{1},
S.~Samsam\,\orcidlink{0000-0001-6311-3801}\auno{1},
M.~Voltolini\,\orcidlink{0000-0002-3843-4854}\auno{4},
M.~Zucali\,\orcidlink{0000-0003-3600-7856}\auno{4}
}

\address{\vspace{0.2cm} $^1$INFN-Milano, Via Celoria 16, 20133, Milan, Italy}
\address{$^2$INFN-Ferrara, Via Saragat 1, 44122, Ferrara, Italy}
\address{$^3$Department of Physics, University of Rome “La Sapienza”, p.le A. Moro, 2-00185 Rome, Italy}
\address{$^4$University of Milan, Dept. Earth Sciences DISTAD, Via Mangiagalli 34, 20133, Milan, Italy}
\address{$^5$University of Milan, Dept. of Physics, Via Celoria 16, 20133, Milan, Italy
\vspace{0.2cm}
\\ Corresponding author: M.~Rossetti~Conti\\
Email address: marcello.rossetti@mi.infn.it}

\begin{abstract}
In this paper we extend further a previous analysis (Ref. \cite{Remotesensing}) of a newly conceived technique applying active interrogation of the tectonic stress evolution in regions hosting active seismogenic faults: the aim is to monitor and detect stable and reliable precursors signals on an adequate time scale, well before the earthquake event, that can play a crucial role in activate alarms to the civil protection system. The precursor signal is based on a continuous measurement of the time evolution of the tectonic stress, realized by interrogating underground, with a high energy collimated muon beam, the piezoelectric fabrics present in granite-like rocks rich in quartz, around a known seismogenic fault region in the Earth's crust. The beam propagation through the rock, in the region across the active fault, brings to the detector, located at the exit of the rock traversal, informations about the amplitude of the piezoelectric field, which in turns scales like the tectonic stress applied to the quartz crystals embedded in the rock. The system named ERMES (Earthquake Reconnaissance using Muon beam Evolution in Silicon dioxide), instead of detecting electro-magnetic waves/signals generated by the piezoelectricity and propagating outside the Earth's crust, as foreseen in other techniques presently under study, it actually probes the piezoelectricity effects from inside the source of the associated electro-magnetic field. It actually probes the near field inside quartz crystals instead of the far field in open space outside the Earth's crust. Here we present a more focused analysis on the manipulation of the muon beam after the rock layer traversal, before reaching the beam detector, using a newly conceived muonic lens, and we explore the maximum capability of rock penetration by the high energy muon beam, in the range of about 3 kilometers of rock thickness for 10 TeV muon beam energy.  Because of the peculiarity of the scenario analysed, in terms of muon beam propagation in such a very thick (km long) target, we carried out cross checks of our previous montecarlo simulations, performed with Fluka, using also \textsc{Geant4}to better clarify the role of secondary muons generated in the interaction of primary muons with the matter in such a long propagation length in solid matter.
\end{abstract}

\maketitle


\section*{Introduction}

Several publications can be found in the literature \cite{ref2_wang2021,ref3_yoshida1998,ref4_sasai1991,ref5_bahari2025,ref6_martinelli2020,ref7_bishop1981} discussing the electro-magnetic fields generated by piezoelectric effects driven by tectonic stress acting in regions of the Earth's crust hosting active seismogenic faults, either in the early proximity of earthquakes or during and soon after them. Piezoelectricity in rocks is associated to the presence of quartz crystals, where its abundance is quite large (up to 30\%) (e.g. granite-like): since the piezoelectric field amplitude scales linearly with the pressure applied to quartz crystals, which in turn depends on the active tectonic stress that varies in time, a direct measurement of this piezoelectric field amplitude can be used for a continuous monitoring of the temporal evolution of the tectonic stress, in particular its up escalation towards the rupture limit, whose overtaking gives rise to the seismic event.

The novelty of the ERMES technique, firstly described in a previous paper \cite{Remotesensing}, and rediscussed here in further details, lies in performing an active interrogation of the sparse piezoelectric fabrics hosted in the rock, instead of carrying out a passive detection of the electro-magnetic signals generated inside the Earth's crust and propagating outside in free space over long distances and with very long characteristic wavelengths. In \autoref{fig:Fluence3km} of Ref. \cite{ref2_wang2021} the author clearly shows the scenario typical of an active fault where rocks rich in quartz host the embedded piezoelectric fabrics: an array of dipoles represents the e.m. source driven by piezoelectricity. A more complete classification of different piezoelectric scenarios is discussed and illustrated in Ref. \cite{ref7_bishop1981} (see figure 4), where several possible spatial and crystallographic distributions of the piezoelectric fabric are shown, with explicit indications of a possible regime of the so-called coherent or true piezoelectricity that can occur in highy strained rocks (e.g., mylonitic). 

In our ERMES analysis, we consider the least favorable scenario of totally incoherent piezoelectricity, where the symmetry axis of the quartz crystals is randomly oriented in space. This leads to a random walk process in the transverse component of the muon momentum crossing the rock layer, driven by the kicks applied by each quartz crystal's piezoelectric field. It is such a probe of the internal piezoelectric field in quartz crystals that brings information to a muon detector, located at the exit of the rock layer traversal, about the piezoelectric field amplitude, hence about the acting tectonic stress at the time of the muon beam crossing (which typically lasts a few micro-seconds for relativistic muons crossing a km-long rock layer). The theoretical model illustrated in Ref. \cite{Remotesensing} derives an expected value for the relative amplitude of the signal carried to the muon detector, represented by the dimensionless quantity $\Delta$, which measures the relative increase of the muon beam spot size 

\begin{equation}
\Delta = \frac{\sigma_{\text{\tiny{TOT}}}}{\sigma_{\text{\tiny{MCS}}}} - 1 = 1.15 \cdot 10^{-4} E_p^2 L_c
\end{equation}

in presence of a piezoelectric-induced random walk process (PRW) as a function of the quartz crystal average length $L_c$ (in m) and the piezoelectric field amplitude $E_p$ (in MV/m), considering that $E_p$ scales linearly with the applied tectonic pressure at 5 MV/m per kbar. In Eq.1 $\sigma_{\text{\tiny{MCS}}}$ represents the Root Mean Square (RMS) beam spot size in absence of PRW (i.e. considering only Multiple Coulomb Scattering (MCS)), while $\sigma_{\text{\tiny{tot}}}$ is the spot size generated by the cumulative MCS+PRW effects.

$\Delta$ achieves values in the range $6.0 \times 10^{-5}$ - $1.4 \times 10^{-3}$ with typical pressures in the range 1-5 kbar and cm-sized quartz crystals. The main technical challenge of an ERMES system is actually to achieve a stable and reliable measurement of $\Delta$ and its time evolution during the time scale of the tectonic stress evolution towards the earthquake.

Therefore, the active interrogation performed by an ERMES system has the big advantage of a direct probe of the inner piezoelectric field inside the fabrics, while a passive detection in open space of the e.m. fields generated by piezoelectricity involves e.m. wave propagation from the fault region towards an external antenna. The big limitation of such a scheme of passive detection is the strong damping of e.m. fields generated in the far field by stochastically arranged arrays of electric dipoles, combined with the very long wavelength associated to these e.m. signals (thousands of km).

According to Ref. \cite{ref2_wang2021}, the time scale to be monitored is, during the final critical phase of the tectonic stress build-up, in the range of a few tens of minutes. Figure 3 of Ref. \cite{ref2_wang2021} illustrates the typical time scale: the critical time interval is the one of plastic loading preceding the rupture/slip event. The evaluated time scale is tens of minutes, with duration scaling with the earthquake magnitude: this implies the capability of an ERMES system to take a single measurement of the tectonic stress within a minute time span (say 100 seconds) at intervals of 1-2 minutes, so to be able to follow the time evolution and measure the time interval in order to achieve also an indication about the incoming earthquake magnitude. As previously discussed (see Ref. \cite{Remotesensing}), if a single measurement requires about $10^{10}$ muons collected on the detector, and the maximum time duration of the measurement must be shorter than 100 seconds, the muon beam average current must reach a capability to deliver up to $10^8$ muons per second in the beam.


\section{Maximum penetrability of high energy muons in granite-like rocks for ERMES systems}
\label{sec:1-MaxPenet}

One of the demands from seismologists is about the maximum achievable capability to investigate large underground areas of the Earth's crust by penetrating km-long thick layers of rock across active seismic faults. 

Previous calculations \cite{Remotesensing} showed that 1 TeV muons can penetrate up to 1 km of granite-like rock, reaching a detector located at the exit with a remaining kinetic energy in excess of 100 GeV, a value quite suitable for efficient detection of the survived muons. 
\begin{figure}[!htbp]
\centering
\includegraphics[width=0.5\textwidth]{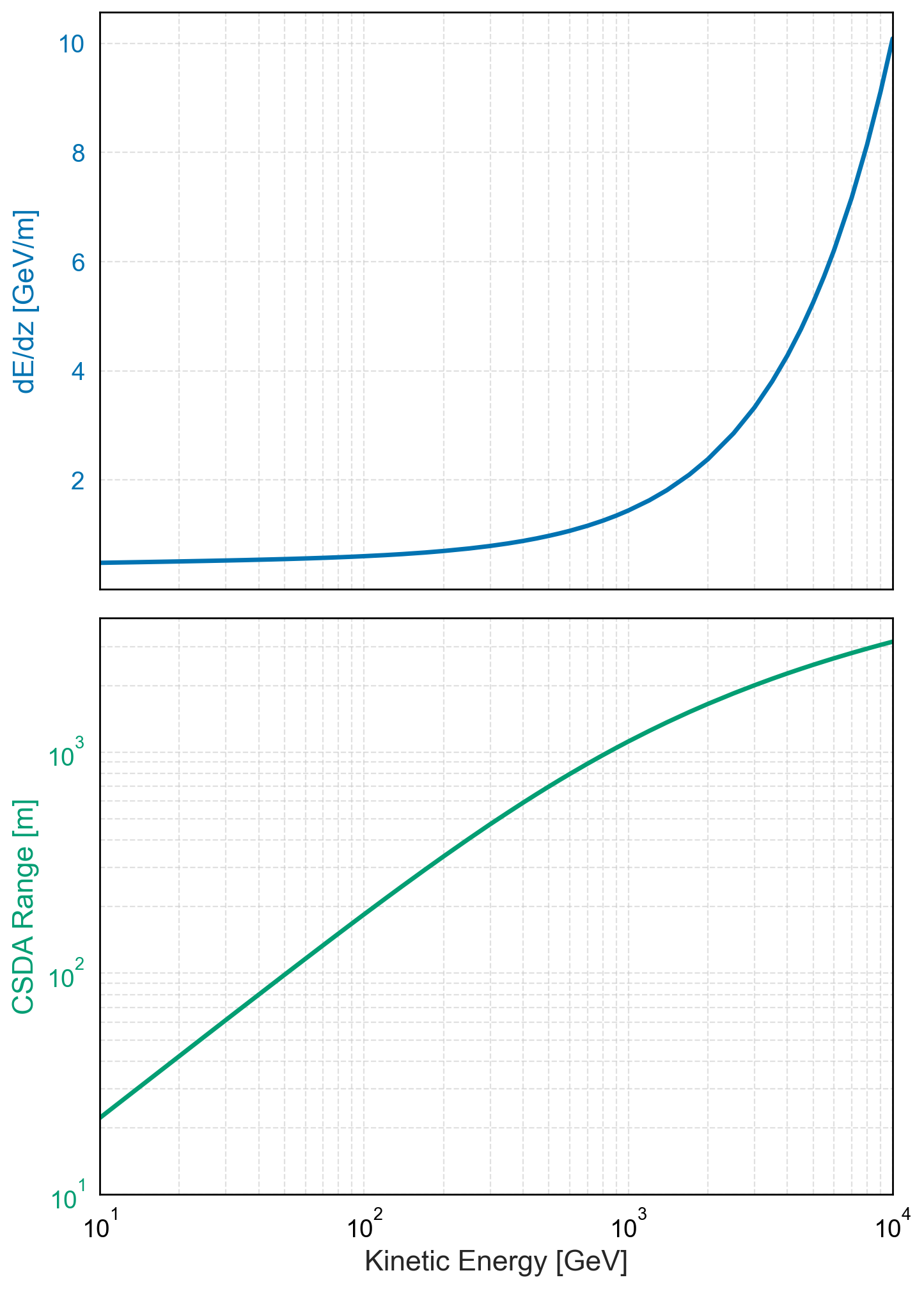}
\caption{Muon energy loss and range in silicon dioxide (fused quartz, SiO$_2$) as a function of the muon kinetic energy $T$ (in GeV, logarithmic scale). 
\textbf{Top panel:} differential energy loss $dE/dz$, expressed in GeV per meter of propagation. 
\textbf{Bottom panel:} CSDA (Continuous Slowing Down Approximation) range in meters, representing the mean path length traveled by a muon before stopping in the material. 
Both quantities are obtained from data reported in \cite{ParticleDataGroup:2024cfk}.}
\label{fig:energy_loss}
\end{figure}

In order to estimate a practical maximum rock thickness we report in \autoref{fig:energy_loss} (upper diagram) the specific energy loss of high energy muons in pure SiO$_2$, together with the average effective range (lower diagram). 

Assuming a maximum muon energy of 10 TeV, which is the maximum energy considered in present studies on muon colliders \cite{collider}, we see that estimated effective range is about 3 km. 
Indeed the specific energy loss clearly shows a behavior of large increase above 2 TeV, as is well known that above this energy, for the material under consideration (amorphous quartz) the energy loss of the muons becomes dominated by bremsstrahlung, instead of ionization, so that the effective range tends to saturate asymptotically and there is no practical gain in increasing the muon energy above 10 TeV in order to increase its penetrability.
Therefore we report in the following the result of a montecarlo simulation of 10 TeV muons crossing 3 km thick amorphous quartz (SiO$_2$) carried out using the code Fluka \cite{Ahdida2022Fluka}. The aim is to evaluate the muon beam energy spectrum evolution through the rock, as well as its angular and transverse spread caused by the interaction with matter, mainly driven by MCS. 

The muon survival ratio is a further important parameter to be monitored. In order to check the estimated results shown in \autoref{fig:energy_loss}, which are based on tabulated data from Ref \cite{ParticleDataGroup:2024cfk}, as well as to quantify the beam spread caused by the 3 km long rock traversal, we ran the montecarlo code Fluka with initial 10 TeV monochromatic muons forming a beam with transverse distribution as plotted in \autoref{fig:SpotSize0-3km} (left diagram, RMS $\sigma_x = 24$ cm). 

\begin{figure}[!htbp]
    \centering
    \includegraphics[width=0.5\linewidth]{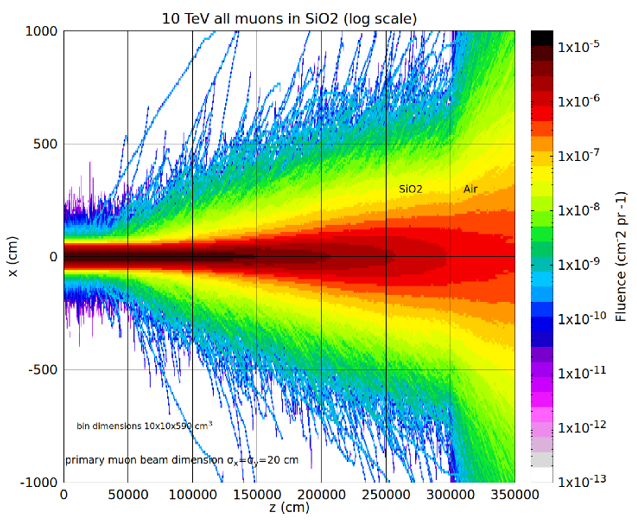}
    \caption{Muon fluence for an initial 10 TeV muon beam, shown on a logarithmic color scale, as a function of propagation distance through a composite medium. The muons propagate first through 3 km of silicon dioxide (SiO$_2$) and subsequently through 500 m of air.}
    \label{fig:Fluence3km}
\end{figure}
Carrying out 10 independent runs with $10^4$ particles each, applying a cut-off for all interactions at 10 MeV, we found that nearly 48 percent of the muons survive the rock traversal and their fluence along the propagation is plotted in \autoref{fig:Fluence3km}, using false colors in logarithmic scale. 
The core of the muon beam stays quite well collimated through the first km, while after that the scattering and probably the recoil due to bremsstrahlung emission of hard photons start to dilute the transverse phase space of the beam, causing a significant de-focusing. 
Nevertheless the beam transverse distribution at the exit ($z=3$ km) is still dense (see right diagram in \autoref{fig:SpotSize0-3km}). 

\begin{figure}[!htbp]
    \centering
    \includegraphics[width=0.7\linewidth]{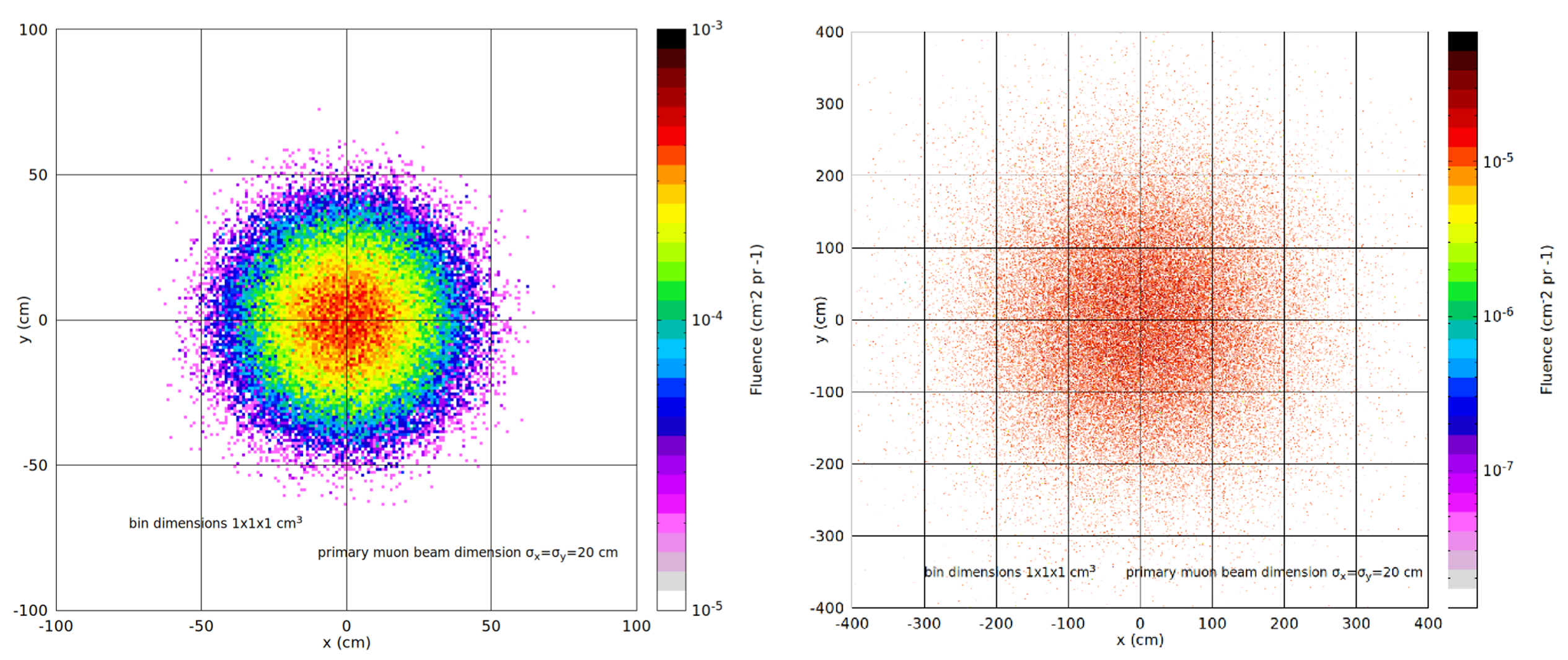}
    \caption{Particle density distribution on the transverse plane at beam entrance (left panel) and after 3 km of propagation (right panel) in SiO$_2$. The density is shown on a logarithmic color scale.}

    \label{fig:SpotSize0-3km}
\end{figure}
\begin{figure}[!htbp]
    \centering
    \includegraphics[width=0.55\linewidth]{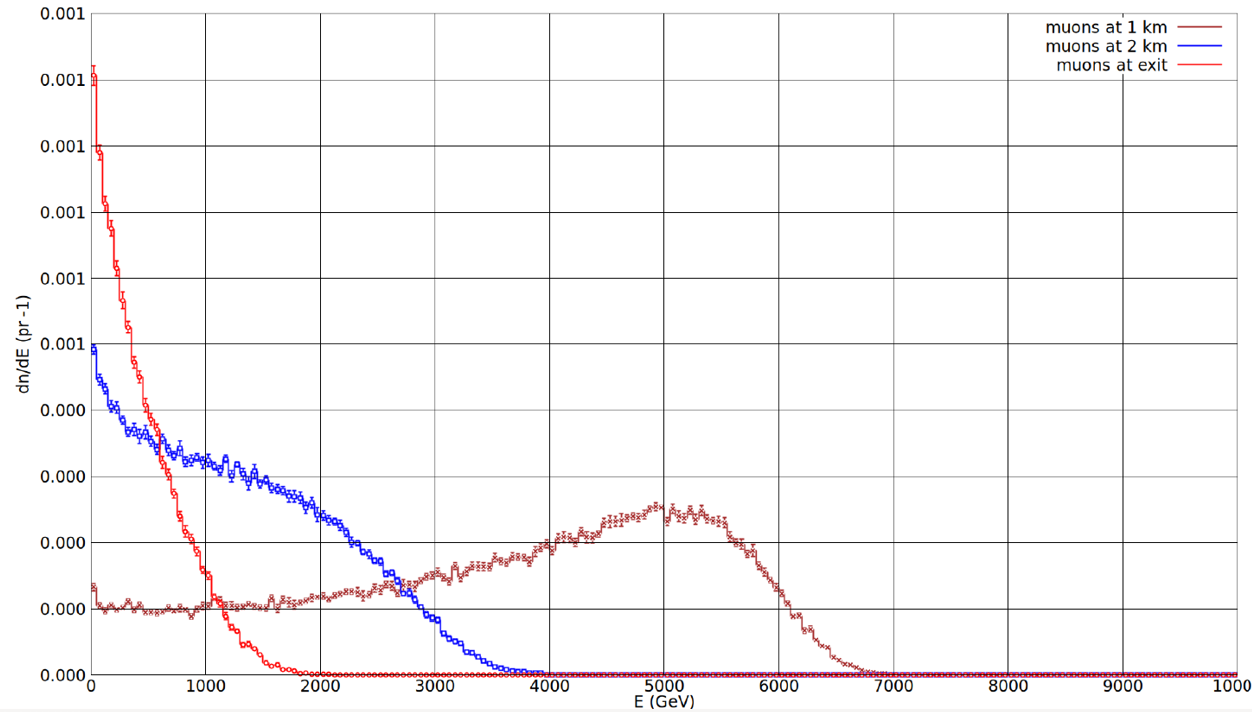}
    \caption{Energy distribution of the muon beam at different propagation depths in SiO$_2$. The spectra are shown after 1 km, 2 km, and 3 km of propagation, illustrating the progressive energy loss and spectral deformation of the beam as a function of depth in the material.}

    \label{fig:Spectra1-2-3km}
\end{figure}

The energy spectra of the muon beam are plotted in \autoref{fig:Spectra1-2-3km} after 1 km of rock thickness, after 2 km, and at the exit (z=3 km): the initial energy spectra (at $z=0$) is a delta-line peaked at 10 TeV. 
The surviving muons still have a relevant energy, on average 500 GeV, well suitable for measurement at a detector downstream the rock exit location. 

These results show, quite well in agreement with tabulated data, that is possible to cross a few km of rock layer with 10 TeV, with a maximum conceivable value probably located between 3 and 4 km, depending on the minimum energy needed for effective muon detection aiming at measuring the effect of the $\Delta$ parameter. 
Further studies are of course needed to assess the performances of silicon strip detectors with high spatial and angular resolution in achieving the requested sensitivity on the expected small values of $\Delta$. 

\section{Geant4 simulations of high energy muon propagation in 600-meters-thick silicon dioxide}
\label{sec:2-Geant4}


In this Section we report numerical results obtained with the montecarlo code \textsc{Geant4} \cite{agostinelli2003geant4,allison2006geant4,allison2016recent}, applied to a previously studied paradigmatic case \cite{Remotesensing}, 500 GeV muon beam launched towards 600 m of SiO$_2$.
The characteristics of the surviving muon beam will be used in the next Section to study the post-focusing effect applied by the muonic lens.

In our study we used the most recent \textsc{Geant4} version (11.4), taking advantage of \textit{FTFP\_BERT} physics list (for more information, see the \textsc{Geant4} documentation \cite{geant4_ftfp_bert}) to consider in our simulation both electromagnetic and hadronic processes. 
This physics list, which is recommended for study of high energy physics, simulates hadronic interactions via elastic, inelastic, and capture processes. The implementation of each process relies on defined sets of cross sections and specific interaction models. 
For electromagnetic interactions, this physics list utilizes the standard \textsc{Geant4} models provided by the \textit{G4EmStandardPhysics} constructor. Among these models there are all the processes for photons and Bremsstrahlung, ionization, single and multiple Coulomb scattering for charged particles.
\begin{figure}[!htbp]
    \centering
    \includegraphics[width=0.6\linewidth]{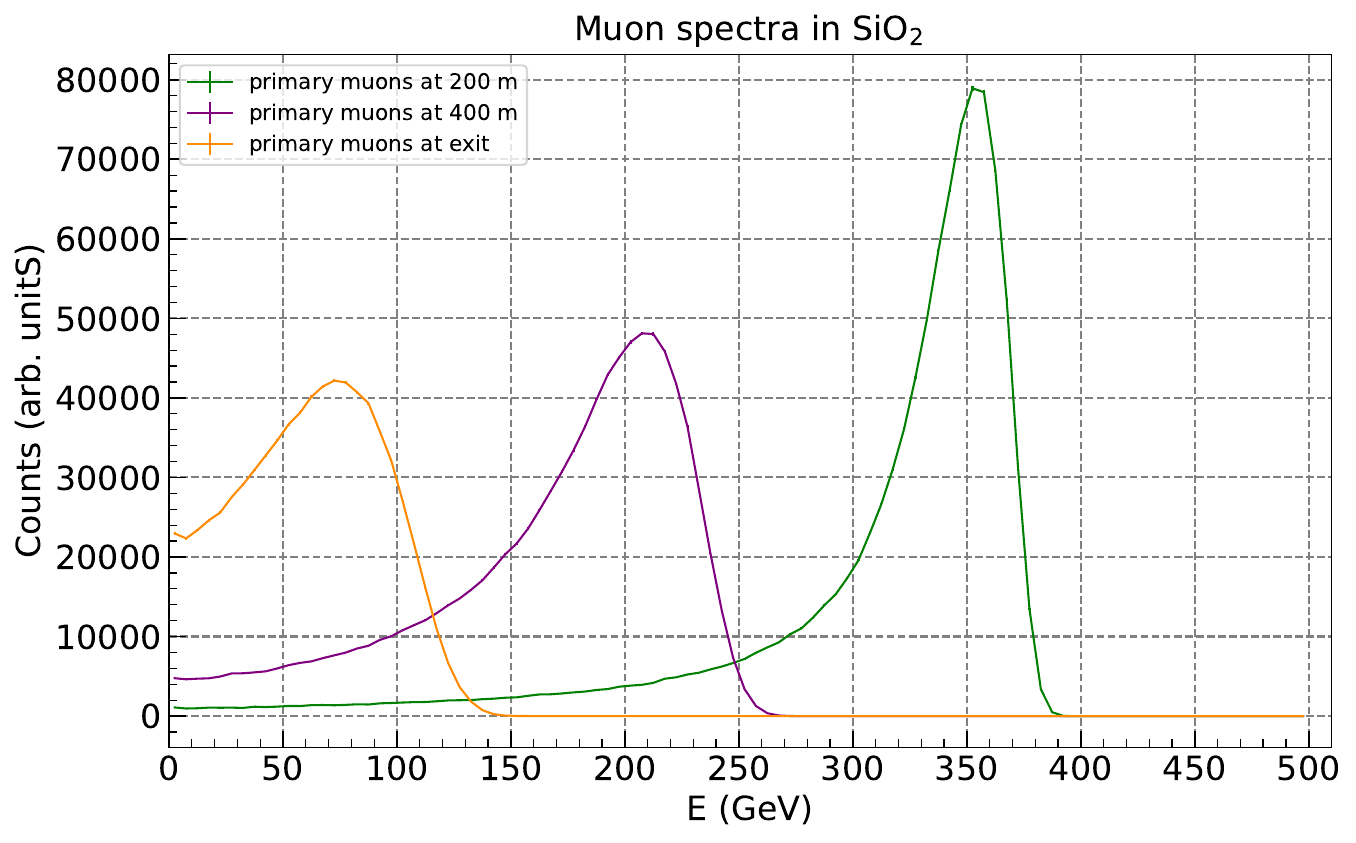}
    \caption{Energy distribution of primary muons at various depths inside SiO$_2$. It is worth noting that the error bars are present, but they are barely visible, since the since the uncertainties on simulated data are very small.}
    \label{fig:primaryMuonsEnergyDistrib}
\end{figure}

As mentioned above, the primary beam was a parallel beam of 10$^6$ muons with an energy of 500 GeV and an RMS size of 25 cm in both transverse directions (x,y). For the secondary particles a production cut of 10 cm was set. We scored the features of the survived particles at various penetration depths (z) inside the rock volume modeled through a 100$\times$100$\times$600 m$^3$ box of SiO$_2$.

\begin{figure}[!htbp]
    \centering
    \includegraphics[width=0.5\linewidth]{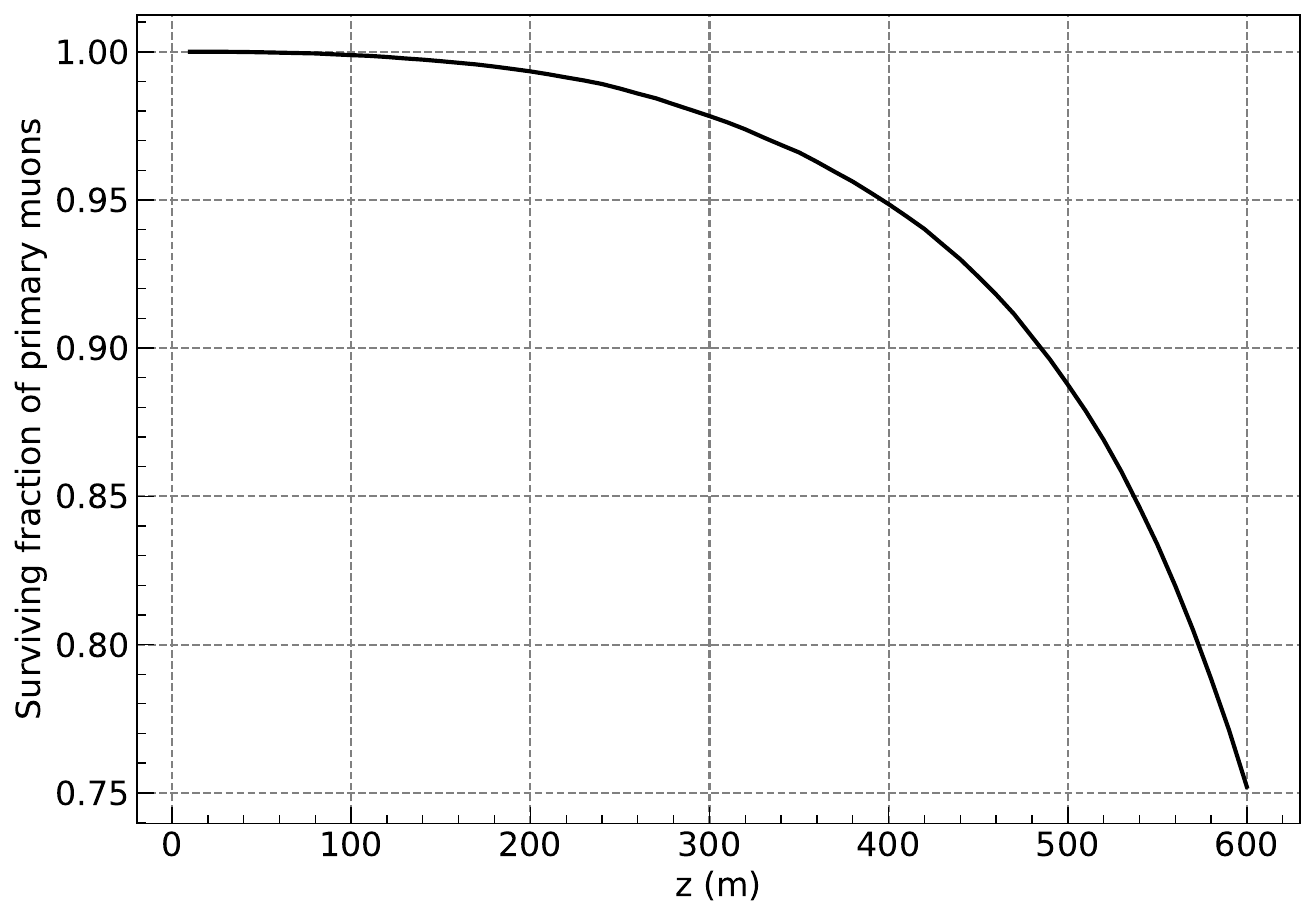}
    \caption{Surviving fraction of primary muons as the penetrate in SiO$_2$ volume.}
    \label{fig:survivingFraction}
\end{figure}

\begin{figure}[!htbp]
    \centering
    \includegraphics[width=0.5\linewidth]{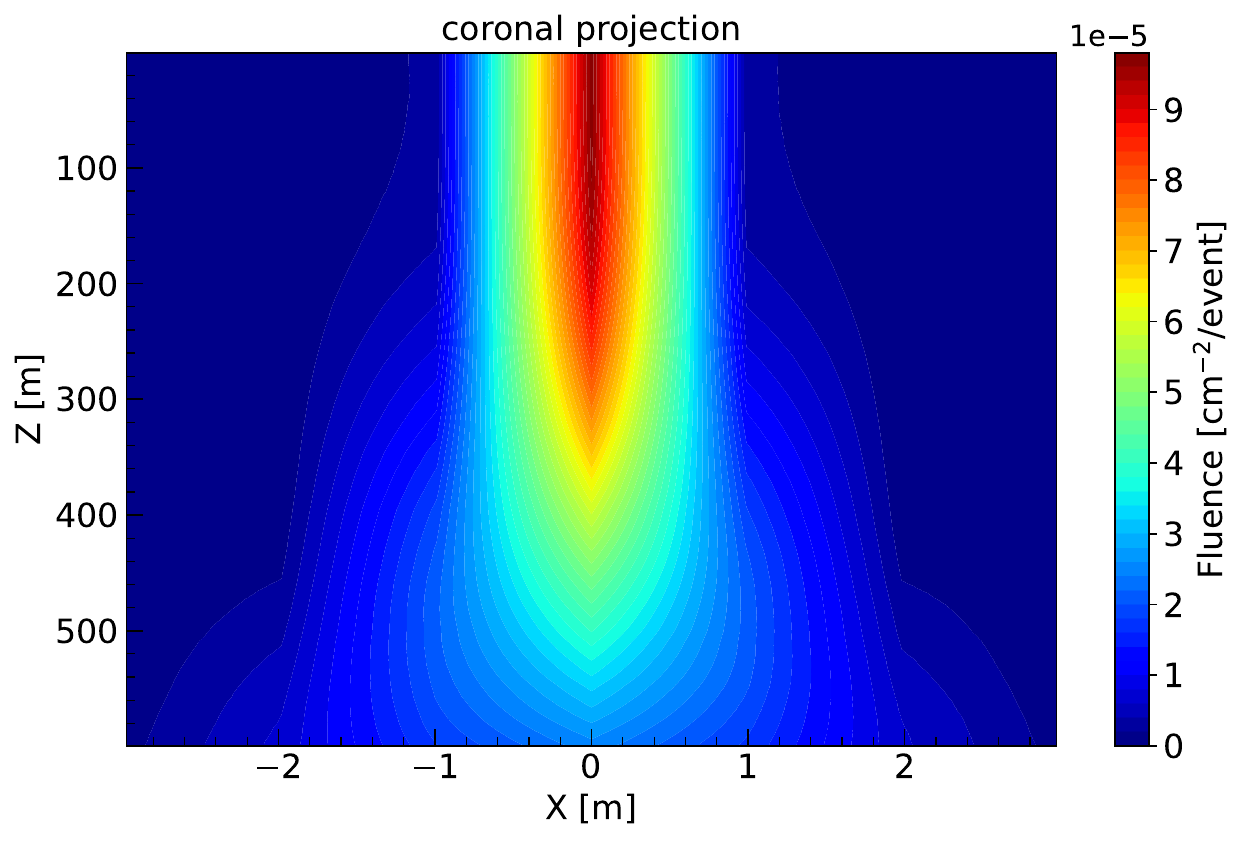}
    \caption{Muon fluence for an initial 500 GeV muon beam as a function of propagation distance through SiO$_2$.}
    \label{fig:muonFluence}
\end{figure}

\begin{figure}[!htbp]
    \centering
    \includegraphics[width=0.7\linewidth]{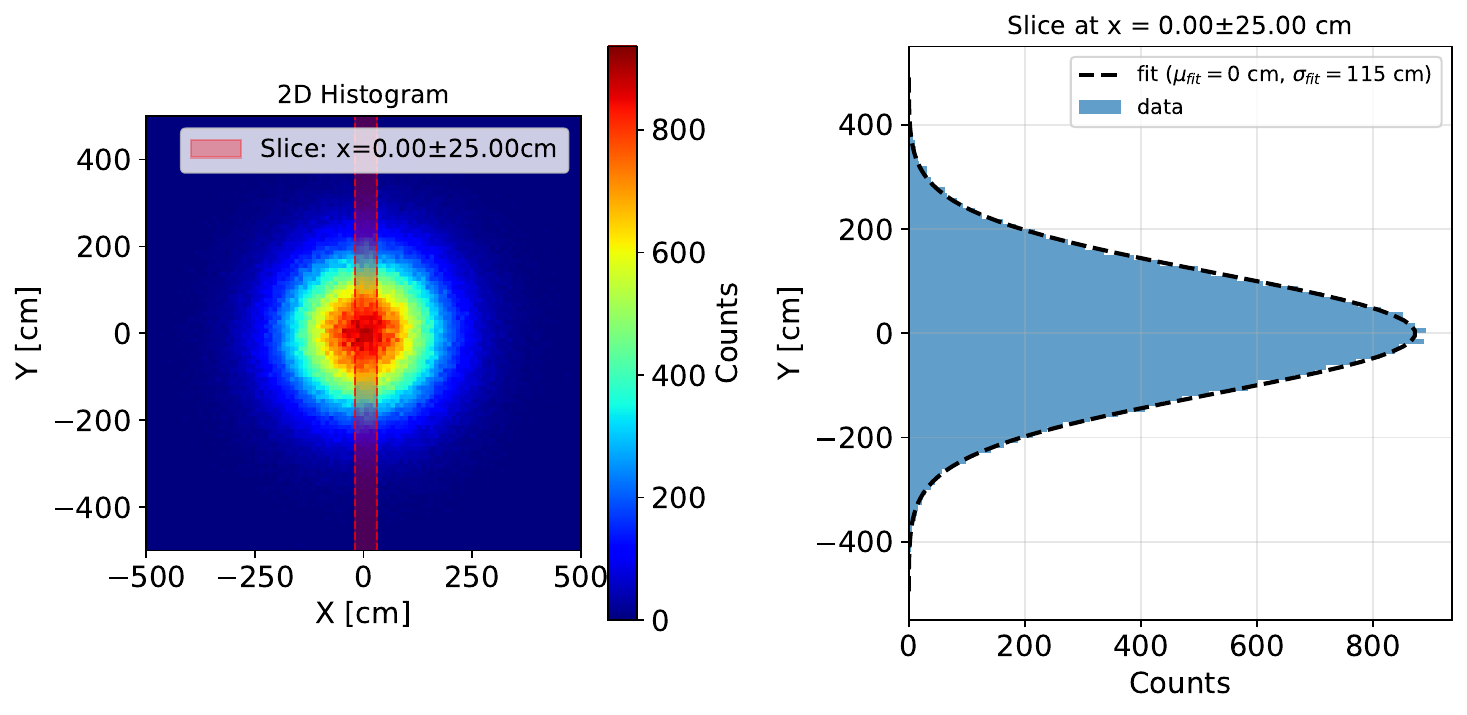}
    \caption{Transverse distribution of the muon beam at the exit surface of the SiO$_2$ volume.}
    \label{fig:exitSpatialDistrib}
\end{figure}

\autoref{fig:primaryMuonsEnergyDistrib} shows the energy spectrum of the primary muons at three different depths inside the considered volume. It is apparent that the primary muons loose about 0.7 GeV/m, in agreement with the slowly increasing part of the $dE/dz$ curve shown in \autoref{fig:energy_loss}, which is dominated by ionization losses. It is also apparent that, as expected, the spread of the spectra increases with the penetration depth, passing form 19\% to 43\%.
The primary muons surviving fraction is shown in \autoref{fig:survivingFraction}. It turns out that 75\% of the impinging muons reach the exit surface. 
The fluence of the muon beam as it propagates through the rock volume and its cross section at exit are shown in \autoref{fig:muonFluence} and \autoref{fig:exitSpatialDistrib}, respectively. 

\begin{figure}[!htbp]
    \centering
    \includegraphics[width=0.5\linewidth]{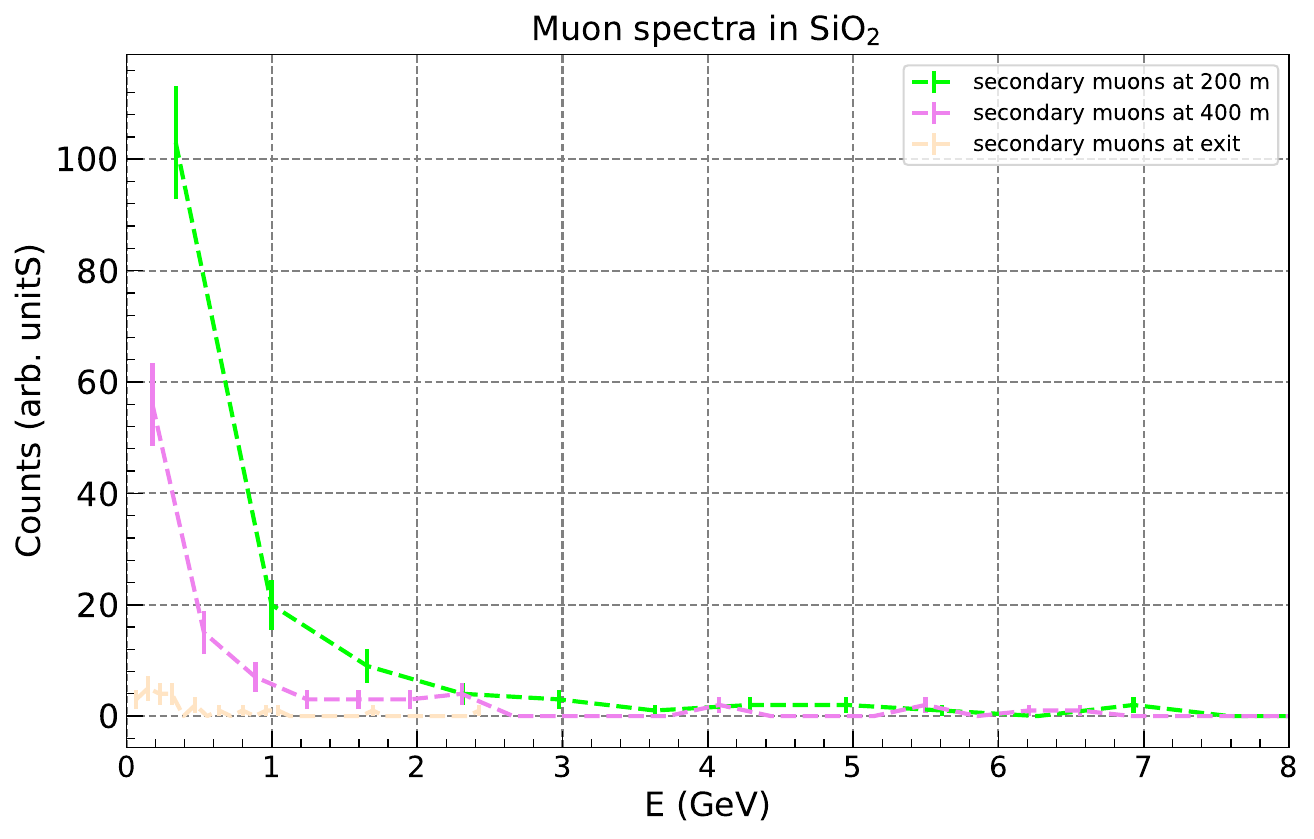}
    \caption{Energy distribution of secondary muons at various depths inside SiO$_2$.}
    \label{fig:secondaryMuonsEnergyDistrib}
\end{figure}

As a result of primary muons interactions, a high numbers of secondary particles are produced, the largest part of which are photons, e$^{\pm}$, neutrons and neutrinos. Only a very small percentage of secondaries are muons, their spectra are shown in \autoref{fig:secondaryMuonsEnergyDistrib}.

From \autoref{fig:secondaryMuonsEnergyDistrib} is apparent that a very small number of secondary muons (0.02\% of the impinging primaries) survives after the first 200 m of rock and their spectrum is bremsstrahlung-like. At the exit, after 600 m of rock the numbers of surviving secondary muons is negligible.

Comparing these results with the ones reported in our previous work \cite{Remotesensing}, it turns out that, apart from a shift on the peak position, the spectra of the primary muons are similar. The peak shift may be due to the fact that in \textsc{Fluka} we neglected the interactions of the muons with the nuclei of the material they passed through. Indeed, the position of the peaks was at a higher energy compared to the results obtained with \textsc{Geant4}, where, instead, we took into account nuclear interactions. Regarding the the spectra of the secondary muons, the discrepancy of the plots in \autoref{fig:secondaryMuonsEnergyDistrib} and those in figure 15 of our previous work, is due to a mislabeling of secondary muons. In \textsc{Geant4} simulations, secondary particles are those that are generated as a result of the interactions of primary particles with the medium they pass through. Conversely, in \textsc{Fluka}, the secondary particles were identified considering some conditions on the scattering angles. In particular, they were actually primaries scattered more than a threshold angle. For these reason they contribute significantly to the total spectrum, which show the same behavior of the spectrum of the muons labeled as primary, namely those little scattered.

The results described above, obtained using \textsc{Geant4}, were benchmarked against the new version of the \textsc{MuAEGIS} (Muon Underground Active Earthquake Genesis Investigation Software) simulation code~\cite{Remotesensing}. 
With respect to our previous work, the code has been significantly improved, primarily in the treatment of energy loss, in order to properly account for transverse cooling effects arising from the deceleration of particles with non-negligible transverse momentum, as is the case in the present study.

The updated energy-loss routine exhibits a pronounced transverse envelope containment, halving the envelope dimension estimated in the previous studies.

\begin{figure}[!htbp]
    \centering
    \includegraphics[width=0.6\textwidth]{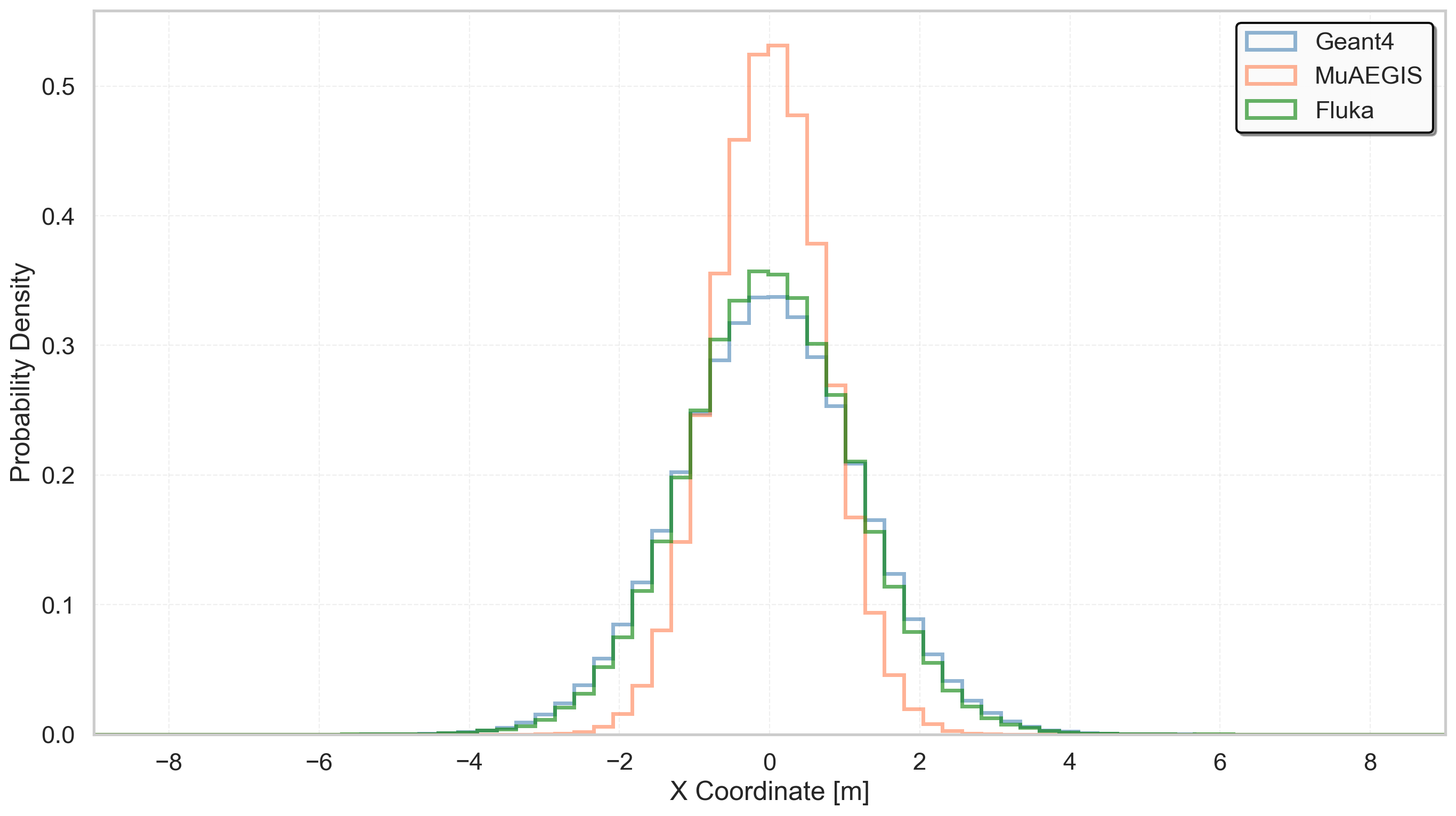}
    \caption{Normalized probability density distributions of the x coordinate for \textsc{Geant4} 
    (blue), \textsc{MuAEGIS} (orange), and \textsc{Fluka} (green) simulations. The histograms are displayed 
    with step-wise contours for improved clarity. The dashed vertical lines indicate the RMS) values for each dataset.}
    \label{fig:x_coordinate_comparison}
\end{figure}

A comparison of the beam transverse extension at the exit of the rock layer obtained by the three codes is shown in \autoref{fig:x_coordinate_comparison}.
The values obtained with \textsc{MuAEGIS} are lower than those predicted by both \textsc{Geant4} and \textsc{Fluka}. 
We attribute this discrepancy, at least in part, to the fact that \textsc{MuAEGIS} propagates effectively monochromatic beams, as it applies an average energy loss that accounts for all interaction processes in SiO$_2$. 
As a consequence, the low-energy tail of the bremsstrahlung spectrum, which would be expected to experience significantly enhanced transverse scattering due to MCS, does not explicitly emerge in the simulation.

In any case, a detailed investigation of the origin of this discrepancy is beyond the scope of the present work and will be the subject of future studies.

\section{Focusing muon beams with large transverse emittance and spot sizes using muonic lenses}
\label{sec:3-MuonLens}
In this Section we describe a conceptual study of a focusing device for the surviving muon beam at the rock layer exit, before transporting the beam to a detector. The surviving beam has a very large beam spot size, of the order of 0.7 meter transverse RMS $\sigma_x$ (corresponding to $\sigma_r\simeq 1$ m for a cylindrical beam, where $\sigma_r$ represents the RMS of the radial coordinate of the particles), as well as a large RMS divergence (10 to 25 mrad), as shown by the simulation results reported in previous Sections. This poses severe demands to any detector considered for the muon beam measurement and characterization, implying that a 6 m diameter round detector should be used to collect the transverse beam density distribution up to at least 6 $\sigma_r$. Focusing the muon beam with a de-magnification of a factor at least 10 would be highly desirable, so to limit its RMS radial spot size $\sigma_r$ to max 10-20 cm, and to keep at the same time its envelope under control so to transport it effectively for some length down to the detector location. Conventional focusing devices like quadrupoles and solenoids are obviously not viable neither applicable to this kind of beams, due to the extremely large inner bore requested. The large kinetic energy of the surviving muons (in excess of 10 GeV) also poses a further challenge to any focusing device in terms of requested focusing gradient. A possible solution is to let the muons propagate through solid matter constituting the focusing device instead of the conventional vacuum pipe, typically used in particle accelerators, that goes through the inner bore of the focusing magnet. At the end, the surviving muons already traveled and propagated through km-long solid matter, a further path of a few meters in length wouldn't change dramatically their beam properties. First of all we need to consider that the power deposition in solid matter by the surviving muons is very small, as is the beam power carried by the surviving muon beam. As a matter of fact a 50 GeV muon beam carrying $10^8$ muon per second has a beam power of just 0.8 W: no threat of damage whatsoever can be expected from such a low beam power. A second important consideration is that a focusing device should desirably maintain the cylindrical symmetry of the muon beam. Third, linear focusing is certainly requested to map the beam transverse distribution without distortions, so to keep unperturbed the effect of the parameter $\Delta$, our precursor signal of the earthquake event at the detector, in particular considering the typical large emittance of the surviving muon beam. 
\begin{figure}[!htbp]
    \centering
    \includegraphics[width=0.35\linewidth]{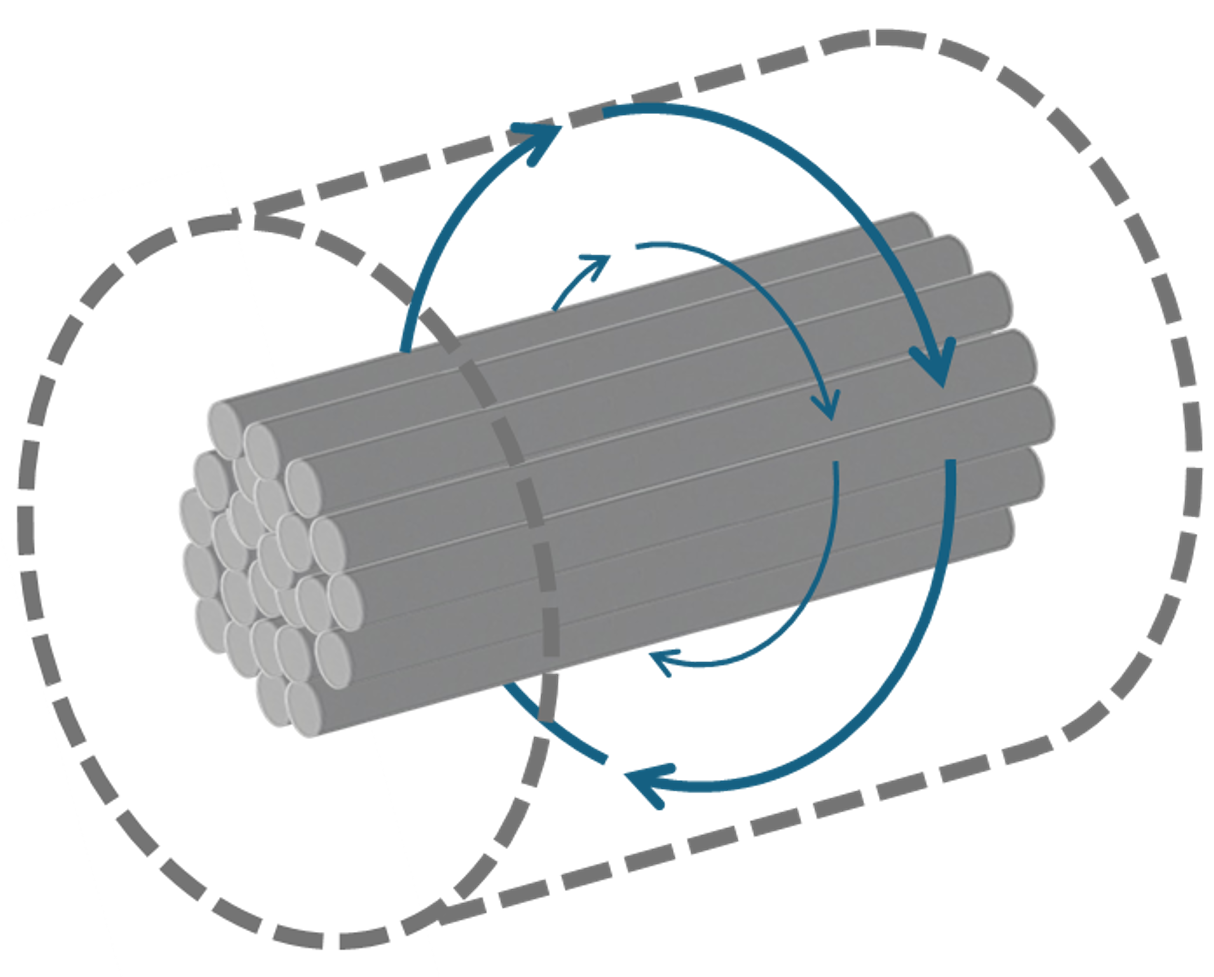}
    \caption{Schematic view of a muonic lens device consisting of parallel aluminum cables filling a cylindrical volume. When a DC current flows through the cables, the device generates a purely azimuthal magnetic field, indicated by blue arrows, which provides focusing of the muon beam.}
    \label{fig:MuonicLens}
\end{figure}
A quite simple device potentially fulfilling all these requests is made of a beam of conducting cables arranged in the form of a cylinder, with diameter as wide as the surviving muon beam at the rock layer exit, carrying a DC electrical current, aligned along the muon beam propagation axis. A wide cylinder mimicking a thick wire carrying an electrical current. Schematically illustrated in \autoref{fig:MuonicLens}, such a device in a simplified 1D approximation will generate an azimuthal magneto-static field ($B_\theta$) linearly increasing from 0 on the symmetry axis up to a maximum value at the edge ($r=R$) of the cylinder equal to $B_\theta = (\mu_0/2) J\,R$ , where $J$ is the current density carried by any wire composing the beam of wires ($\mu_0$ is the magnetic permeability, $\mu_0=4\pi\times10^{-7}$). Considering Aluminum as the conducting material of the wires, since its radiation length and atomic number Z are quite close to those of SiO$_2$, its conventional current density is 2.5 A/mm$^2$ , as is typically used in electrical networks made of aluminum without external cooling. The focal length $F_{oc}$ of such a device, that we may refer to as \textit{muonic lens}, in the thin lens approximation is given, in practical units, by

\begin{equation}
F_{oc} = 5.6\times10^5\,\gamma / (L[m]\,J[A/m^2])
\end{equation}

in the case of relativistic muons (i.e. for $\gamma \gg 1$, with muon rest mass $m_{\mu} = 105.7$ MeV), where $L$ is the total length of the lens and gamma the Lorentz relativistic factor of the muon ( the muon kinetic energy $T_\mu$ is given by $T_\mu = (\gamma-1) m_\mu c^2$). Incidentally we note that the focusing gradient $K_r$ of the muonic lens can be easily derived as  $K_r = e \mu_0 J/(m_\mu c \gamma)$, that in practical units reads ($e$ electron charge, and $c$ speed of light in vacuum):

\begin{equation}
K_r = 1.8\times 10^{-6} J[A/m^2]/\gamma
\end{equation}

A practical example of reference for our case of interest consists of a muonic lens 10 meters long, with a radius $R$ = 3 meters, carrying a current density of 2.5 A/mm$^2$ (max $B_\theta = 4.7$ T at the edge), focusing the surviving muon beam at 50 GeV energy ($\gamma=474.$). Its focal length, estimated with the simple 1D linear model discussed above, would be $F_{oc}$ = 11.2 m.  The specific energy loss of 50 GeV muons in pure Aluminum is about 0.6 GeV/m, implying a nearly 6 GeV average energy loss in the propagation through the muonic lens, which looks at all sustainable by the surviving muons. The average divergence angle accumulated through a 10 m long muonic lens due to MCS is also quite small w.r.t. the beam divergence at the exit of the lens: the former is about 2-3 mrad while the latter is in the range 50-100 mrad, i.e. the ratio between the RMS spot size at entrance and the focal length. Incidentally, this muonic lens dissipates nearly 16 MW of electrical power in omhic losses (power density of about 57 
W/dm$^3$), that must be switched on only during the time intervals of the active interrogation monitoring, during which the muon beam is launched through the rock layer.  
\begin{figure}[!htbp]
    \centering
    \includegraphics[width=0.95\linewidth]{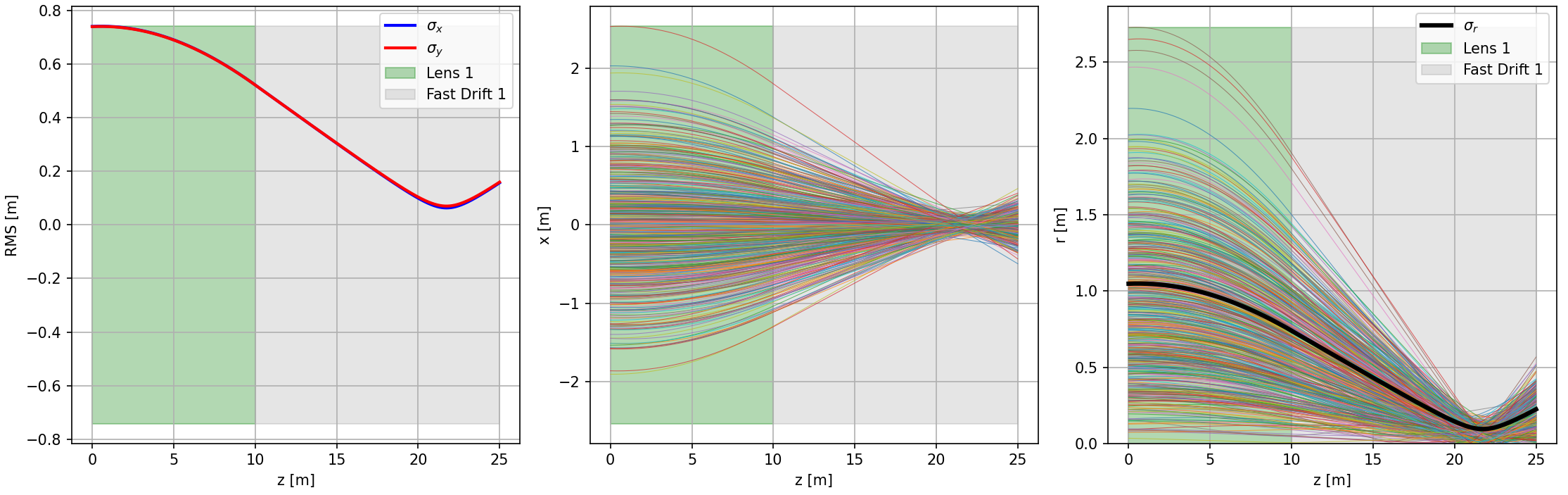}
    \caption{Focusing of a 50~GeV muon beam obtained with a muonic lens of 3~m radius and maximum magnetic field of 3.5~T. 
    \textbf{Left:} evolution of the transverse beam envelope along the longitudinal coordinate $z$, shown in terms of the horizontal and vertical RMS beam sizes. 
    \textbf{Center:} sagittal-plane trajectories of a representative subset of 300 particles, illustrating the collective focusing action of the lens and subsequent drift. 
    \textbf{Right:} radial coordinate of the same particle subset as a function of $z$, together with the corresponding RMS radius. 
    The green shaded region indicates the lens, while the gray shaded region denotes the fast drift section.}
    \label{fig:MuonicFocus}
\end{figure}

We present in \autoref{fig:MuonicFocus} the beam tracking results through a muonic lens of the surviving muon beam, showing a very good focusing effect. 
In order to provide a more quantitative assessment, a 500 GeV muon beam was transported over a distance of 600 m through a medium composed of amorphous SiO$_2$ and quartz crystals. 
The simulation have been described in Ref. \cite{Remotesensing} and includes the effects of MCS and PRW. 
At the exit of the medium, the mean beam energy is reduced to approximately 50 GeV due to the interaction with the medium. 
An energy selection is subsequently applied by rejecting particles whose relative energy deviation exceeds 5\% with respect to the mean beam energy.

The selected beam is then focused by a muonic lens with a length of 10 m and a radius of 3 m, providing a maximum magnetic field at the aperture of $B_{\mathrm{max}} = 3.5$ T. 
The lens generates a focal point located 26.4 m downstream of the SiO$_2$ exit. 
At the focal position, the beam envelope retains cylindrical symmetry and reaches a minimum transverse RMS size of $\sigma_r = 121$ mm.

A real 3D study of the muonic lens will have to be developed in the next future, assessing the edge effects associated to the finite length of the muonic lens, as well as the distortion to the magnetic field linear distribution due to the current inlets at the injection side and the exit side of the lens. 
These fringing fields will definitely spoil the beam quality, in other words they will increase the muon beam emittance: nevertheless, in our case we are dealing with very large emittances of the surviving muon beams, as clearly shown in previous sections the normalized RMS emittance levels reached at the end of the rock traversal, mainly caused by the MCS effect, are up to meters$\,\cdot\,$radians, i.e. $10^6$ mm$\,$mrad (in conventional units), that is 4-5 orders of magnitude larger than the typical emittance values requested by muon colliders \cite{collider}. 
Therefore we expect a tolerable beam quality degradation due to the fringing fields in the muonic lens: from the point of view of the focusing magnetic field spatial distribution, the muonic lens is at all similar to the discharge capillaries used in plasma accelerators to manipulate the electron beam (either focusing or deflecting)\cite{frazzitta2024theory,pompili2018focusing,pompili2024guiding}.

Motivated by the demonstrated effectiveness of the muonic lens in focusing the beam emerging from the Earth's crust, we investigated the feasibility of employing such elements as building blocks of a periodic lattice for long-distance muon beam transport. 
To this end, four focusing and four defocusing muonic lenses were arranged in an alternating sequence, yielding a periodic beam dynamics analogous to a FODO lattice, while preserving the cylindrical symmetry of the beam. 
The main parameters of the focusing and defocusing lenses are summarized in \autoref{tab:muonic_lattice}.

\begin{table}[!htbp]
\centering
\renewcommand{\arraystretch}{1.4}
\setlength{\tabcolsep}{10pt}
\caption{Main parameters of the periodic muonic-lens lattice.}
\label{tab:muonic_lattice}
\begin{tabular}{lcccccccc}
\toprule
 & L1 (F) & L2 (D) & L3 (F) & L4 (D) & L5 (F) & L6 (D) & L7 (F) & L8 (D) \\
\midrule
$r~[\mathrm{m}]$        & 0.8 & 0.4 & 0.8 & 0.4 & 0.8 & 0.4 & 0.8 & 0.4 \\
$L~[\mathrm{m}]$        & 1.0 & 2.3 & 2.0 & 2.3 & 2.0 & 2.3 & 2.0 & 2.3 \\
$B_{\max}~[\mathrm{T}]$ & 0.5 & 0.5 & 0.5 & 0.5 & 0.5 & 0.5 & 0.5 & 0.5 \\
\bottomrule
\end{tabular}
\end{table}

\begin{figure}[!htbp]
    \centering
    \includegraphics[width=0.70\linewidth]{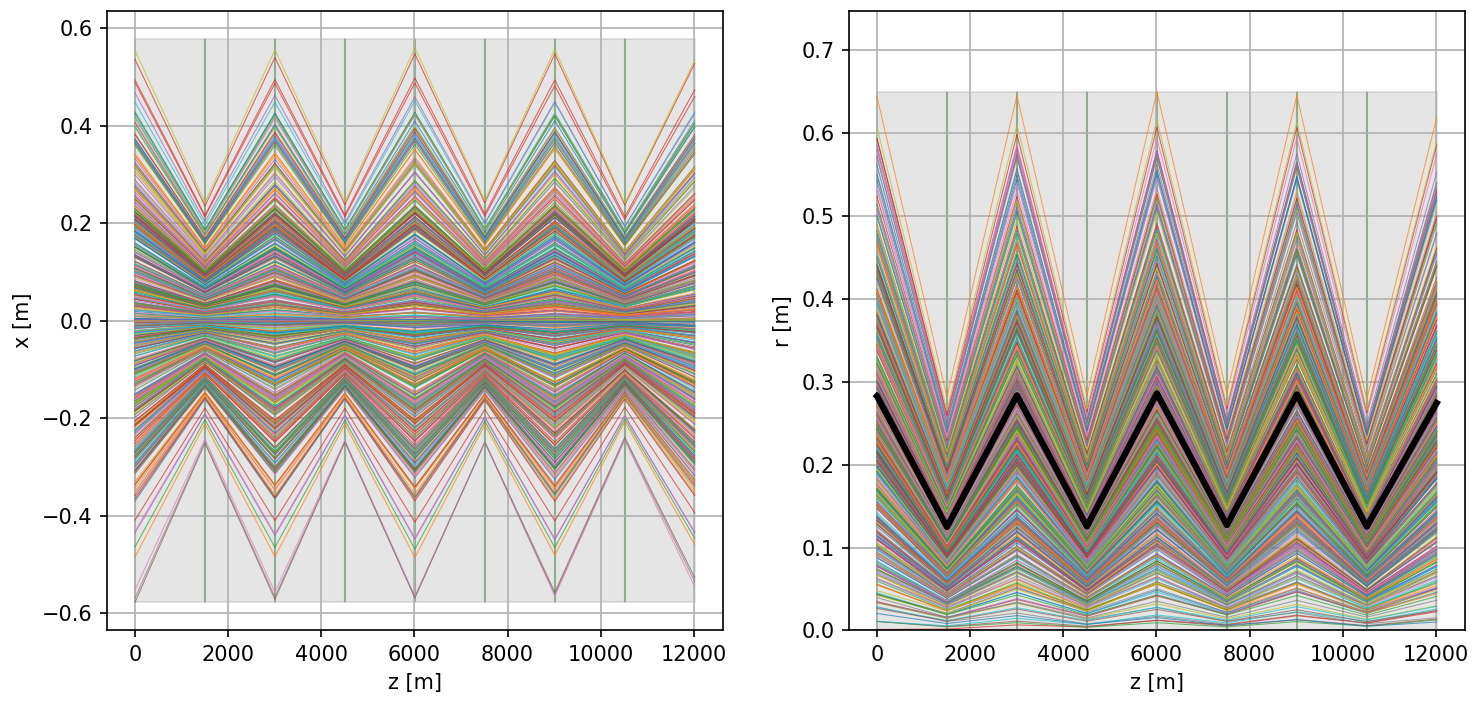}
    \caption{Periodic transport lattice for a 500~GeV muon beam with normalized horizontal emittance 
    $\varepsilon_{n,x}=50~\mathrm{mm\,mrad}$ and transverse RMS beam size 
    $\sigma_{x-y}=20~\mathrm{cm}$. 
    The lattice consists of eight muonic lenses, arranged in an alternating focusing--defocusing (FODO-like) configuration with four focusing and four defocusing elements. 
    \textbf{Left:} sagittal-plane trajectories of a representative subset of 300 particles, illustrating the periodic oscillations along the lattice. 
    \textbf{Right:} radial coordinate of the same particle subset as a function of the longitudinal coordinate $z$, together with the corresponding RMS radius (black curve).}
    \label{fig:MuonicFODO}
\end{figure}

A muon beam with an initial energy of 500 GeV, normalized transverse emittance $\varepsilon_{n,x-y} = 5 \times 10^{3}$ mm mrad, and RMS transverse beam size $\sigma_{x-y} = 20$ cm was successfully transported over a distance of 12 km. 
The tracking of a representative subset of transported particles is shown in \autoref{fig:MuonicFODO}, clearly demonstrating the periodic nature of the beam transport.

\section{Conclusion}
\label{sec:Conclusion}

In this work we have advanced the ERMES concept for active interrogation of tectonic stress using high-energy muon beams propagating through quartz-rich seismogenic regions. Through detailed Monte Carlo studies with \textsc{Fluka} and \textsc{Geant4}, we have shown that muons with energies up to 10 TeV can traverse rock thicknesses of the order of 3 km with significant survival probability and residual energy suitable for detection, while maintaining a measurable phase-space structure despite scattering and radiative losses. The consistency between the two simulation frameworks strengthens confidence in these results and clarifies the negligible role of secondary muons at large depths. Clearly the challenge of these simulations is not in evaluating the single particle (muon) dynamics, but indeed the collective beam phase space evolution in propagating the muon beam through such a thick solid target, as is the km-class rock layer considered here to model the active fault Earth's crust region that must be actively interrogated. Finally, we have introduced the concept of a muonic lens to refocus the large-emittance beams emerging from the rock, which appears essential for enabling practical detection of the ERMES signal. Future work will address detector optimization, realistic geological modeling, and system engineering toward assessing the feasibility of ERMES as a novel earthquake precursor monitoring technique. The aim is to achieve the capability to forecast an earthquake event generated by an active seismogenic fault, not only in a time interval useful for a general awareness concerning the potential risk, but also for an estimation of its expected magnitude. This would be accomplished by a continuous monitoring and surveillance of the growing signal registered in the muon detector by the increasing effects of the applied tectonic stress, which in turns affects the muon beam propagation underground through the rock around the active fault via the PRW (piezoelectric random walk) effect. In a nutshell ``quartz whispers, muons listen''.

\bibliographystyle{unsrt}
\bibliography{bib}
\end{document}